\documentclass{emulateapj}
\usepackage{graphicx}
\usepackage{multirow}

\shorttitle{Global Optimization methods for Gravitational Lens Systems with Regularized Sources}
\shortauthors{Rogers \& Fiege}

\begin{document}
\textheight 9.0in

\title{Global Optimization methods for Gravitational Lens Systems with Regularized Sources}
\author{Adam Rogers and Jason D. Fiege}
\affil{Department of Physics and Astronomy, The University of Manitoba, Winnipeg, Manitoba, R3T-2N2, Canada}
\email{rogers@physics.umanitoba.ca}
\slugcomment{To appear in ApJ, 759, 27, Nov. 1, 2012}

\begin{abstract}
Several approaches exist to model gravitational lens systems. In this study, we apply global optimization methods to find the optimal set of lens parameters using a genetic algorithm. We treat the full optimization procedure as a two-step process: an analytical description of the source plane intensity distribution is used to find an initial approximation to the optimal lens parameters. The second stage of the optimization uses a pixelated source plane with the semilinear method to determine an optimal source. Regularization is handled by means of an iterative method and the generalized cross validation (GCV) and unbiased predictive risk estimator (UPRE) functions that are commonly used in standard image deconvolution problems. This approach simultaneously estimates the optimal regularization parameter and the number of degrees of freedom in the source. Using the GCV and UPRE functions we are able to justify an estimation of the number of source degrees of freedom found in previous work. We test our approach by applying our code to a subset of the lens systems included in the SLACS survey.
\end{abstract}

\section{Introduction}
\label{sec:intro}
Methods for modeling gravitational lens systems are divided into a broad dichotomy between schemes that require a parameterized analytical model for the source intensity distribution, and schemes that assume only a pixelated source with no underlying model.  Methods that parameterize the source intensity distribution are often quite easy to implement, but assume a priori knowledge of the source structure.  Schemes that make use of a pixelated source are generally more complex, but offer greater flexibility since no parametric form is assumed for the source.  This paper makes use of both parameterized and pixelated source models, exploiting the benefits provided by each.

Lens inversion schemes based on analytical source models  assume an intensity distribution $\mbox{$I_s$}(\mbox{\boldmath$\beta$})$ in the source plane $\mbox{\boldmath$\beta$}$. A model of the lens density is then used to calculate a ray-tracing from the image plane $\mbox{\boldmath$\theta$}$ to the source plane using the thin lens equation
\begin{equation}
\mbox{\boldmath$\beta$}(\mbox{\boldmath$\theta$}) = \mbox{\boldmath$\theta$} - \mbox{\boldmath$\alpha$}( \mbox{\boldmath$\theta$} ),
\end{equation}
where $\mbox{\boldmath$\alpha$}(\mbox{\boldmath$\theta$})$ is the deflection angle field calculated from the projected lens potential \citep{schneider85, SEF, wam}. Since gravitational lensing conserves surface brightness \citep{kayserSchramm}, the lensed image intensity is easily found by
\begin{equation}
I(\mbox{\boldmath$\theta$})=I_s(\mbox{\boldmath$\beta$}(\mbox{\boldmath$\theta$}))
\end{equation}
for an assumed parametric source intensity function $I_s$. The resulting lensed image $I(\mbox{\boldmath$\theta$})$ is then convolved with a point spread function (PSF) and compared with the data. The $\chi^2$ statistic is minimized over the combined set of lens and source parameters using non-linear methods for parameter search and global optimization.

S\'{e}rsic profiles \citep{sersic} are widely used for galaxy scale sources, as defined by
\begin{equation}
I_s(r) = I_0 \exp \{-k(n)[(r/r_0)^{-n} -1] \},
\label{sersicEq}
\end{equation}
which assumes intensity $I_0$ at the scale length $r_0$ and shape index $n$. The shape index controls the curvature of the profile, where most galaxies have profiles with $0.5 < n < 10$. The \citet{deV} profile is recovered for $n=4$, and the exponential disk is found by setting $n=1$. The scaling factor $k(n)$ is used to normalize the distribution such that half the total luminosity is within $r_0$.

Due to their flexibility and simple physical interpretation, S\'{e}rsic functions are commonly used to model lensed sources \citep{marshall, SLACSV, brewerBayes}. However, more complicated analytical source functions have also been used to approximate the varied and complex morphologies of galaxies and can include hundreds of parameters in extreme cases \citep{tyson}. In general, analytical models are used because they are typically fast to evaluate and provide an intuitive understanding of the resulting source.

As useful as analytical models are, they may not be flexible enough to describe complex sources and may bias the lens parameters during $\chi^2$ minimization to compensate for the artificial constraints imposed by their assumed analytical form.  Pixelated source models were introduced to move past this limitation. This approach represents the source plane intensity as a set of basis functions, each having an adjustable parameter that represents the surface brightness of the source plane at a given pixel. The semilinear method treats each pixel as a basis function and minimizes the mismatch between model and data by manipulating the brightness of each source pixel $s_j$ independently \citep{WD03, TK04, suyu06}.

The semilinear method divides the lens modeling problem into a non-linear ``outer loop'' problem that solves for lens parameters, and an ``inner loop'' problem that solves for the pixelated source, assuming a fixed set of lens parameters.  An important benefit of this approach is that the inner loop problem is linear and therefore does not require complicated nonlinear optimization routines.
The blurring and lensing effects are expressed by the matrix $\mbox{\boldmath$f$}=\mbox{\boldmath$BL$}$. The lensing matrix $\mbox{\boldmath$L$}$ encodes the ray tracing operation from the image plane to the source plane and forms the lensed image of a given source brightness distribution. In this work we make use of a bilinear interpolation scheme, where the center of each image pixel is traced to a position on the source plane using the lens equation. Then the brightness of an image pixel is found by a weighted average of the four source pixels that enclose each back-traced ray \citep{TK04, koopmans05}. This choice is not unique and many different kinds of interpolation schemes have been studied in the literature including nearest neighbor \citep{WD03}, adaptive source pixel tilings \citep{DW05} and delaunay triangulations \citep{vegetti}. We plan on studying the effects of a variety of such interpolation schemes in future work.

The blurring matrix $\mbox{\boldmath$B$}$ describes the effect of the PSF on the resulting lensed image.  By minimizing the $\chi^2$ statistic with respect to the source plane intensities $s_j$, the least-squares form of the problem is exposed:
\begin{equation}
\mbox{\boldmath$F$}^T \mbox{\boldmath$Fs$} = \mbox{\boldmath$F$}^T \hat{\mbox{\boldmath$d$}},
\label{LS}
\end{equation}
where $\mbox{\boldmath$F$}$ is the lens matrix divided by the errors in the data, $F_{ij}=f_{ij}/ \sigma_{i}$, and $\mbox{\boldmath$s$}$ is a ``flattened'' image vector containing
the intensities of the source plane pixels \citep{WD03, koopmans05}. The vector $\hat{d}_i=d_i/ \sigma_i$ is the data vector $\mbox{\boldmath$d$}$ normalized by the noise $\sigma_i$. This type of problem has been well studied in the context of the standard image deconvolution problem \citep{golubHeathWahba, vogel, regu, nagyRestoreTools}, which seeks to remove the distortion introduced by a blurring function (PSF).

In general, the solution of Equation \ref{LS} requires regularization to stabilize the inversion of the system matrix $\mbox{\boldmath$F$}^T\mbox{\boldmath$F$}$ \citep{koopmans05}. The modified matrix is then given by
\begin{equation}
\mbox{\boldmath$M$}=\mbox{\boldmath$F$}^T\mbox{\boldmath$F$}+\lambda \mbox{\boldmath$H$}^T\mbox{\boldmath$H$},
\label{M}
\end{equation}
where $\mbox{\boldmath$H$}$ is a regularization matrix and $\lambda$ a multiplier that controls the amount of regularization added to the problem. The simplest case, zeroth order regularization, assumes that $\mbox{\boldmath$H$}=\mbox{\boldmath$I$}$. This scheme regularizes the problem by seeking the solution $\mbox{\boldmath$s$}$ that has minimal intensity over the source plane. Higher order regularization schemes are also commonly used, such as curvature regularization that uses the second order derivatives of $\mbox{\boldmath$s$}$ to smooth the solution by minimizing the curvature over the source plane. Regularization schemes seek to impose physicality constraints on the source intensity to select a smoothly varying and physically realistic solution from the many alternatives that exist to solve the ill-posed system.  Linear regularization schemes were studied in depth by \citet{suyu06}.

Following our previous work \citep{rogersFiege2011}, we use the Qubist Optimization Toolbox \citep{qubist} to find the nonlinear lens parameters varied in the outer loop of the lens inversion problem. The Qubist Toolbox contains several non-linear global optimization routines including Ferret, an advanced genetic algorithm (GA), and Locust, a particle swarm optimizer (PSO).  In the inner loop, we solve the least squares problem of the semilinear method using Krylov subspace methods \citep{bjorck}. Krylov subspace methods are well known in the image deblurring community and have been studied in the context of deconvolution problems at length \citep{regu, nagyRestoreTools}. This class of optimization routines include the conjugate gradient method for least squares problems (CGLS) and the steepest descent (SD) method. Krylov methods are attractive because they naturally regularize ill-posed problems and are efficient at solving large scale problems. We previously studied the performance of the GA and PSO methods on test problems using simulated lens data \citep{rogersFiege2011}.  In that work we found that the GA explored the parameter space more thoroughly than the PSO, although the PSO was slightly faster to converge.

In this work, we will explore parameter selection methods to determine an appropriate value for the regularization constant in the semilinear method, and use the Ferret GA with our lens code to model data from the SLACS survey. We use a two stage approach to the lens modeling problem: we begin the optimization with analytical sources to estimate the approximate position of the globally optimal lens parameters, and switch to a pixelated source for further model refinement once the global optimizer has converged.

\section{Gravitational Lens Source Deconvolution}

The semilinear method with regularization describes gravitational lens modeling in the context of a least squares problem, where we seek a vector $\mbox{\boldmath$s$}$ that minimizes
\begin{equation}
g=||\mbox{\boldmath$Fs$}-\hat{\mbox{\boldmath$d$}}||^2+\lambda||\mbox{\boldmath$Hs$}||^2.
\label{minLS}
\end{equation}
The first term in this sum is the $\chi^2$ between the model and observed images, while the second term quantifies the strength of the regularization. In general, gravitational lensing produces multiple images, so $\mbox{\boldmath$F$}$ is a rectangular $N\times M$ matrix ($N > M$), where $N$ is the number of image pixels involved in the inversion and $M$ the number of source pixels.

The most direct method to solve the least squares problem is to decompose $\mbox{\boldmath$F$}$ using the singular value decomposition \citep[SVD;][]{golubSVD},
\begin{equation}
\mbox{\boldmath$F$}=\mbox{\boldmath$U \Sigma V$}^T,
\label{eqF}
\end{equation}
where $\mbox{\boldmath$\Sigma$}$ is an $N \times M$ diagonal matrix composed of a set of non-zero, non-increasing elements $\Sigma_{jj}=\nu_j$ such that $\nu_1 \geq \nu_2 \geq ,..., \geq \nu_M$. These diagonal elements are the singular values of $\mbox{\boldmath$F$}$, defined as the eigenvalues of $\mbox{\boldmath$F$}^T\mbox{\boldmath$F$}$ and $\mbox{\boldmath$FF$}^T$, both of which produce identical sets of non-vanishing eigenvalues. The $\mbox{\boldmath$U$}$ and $\mbox{\boldmath$V$}$ matrices are orthogonal $(N\times N)$ and $(M \times M)$ matrices respectively. We denote the columns of these matrices as $\mbox{\boldmath$u$}_i$ and $\mbox{\boldmath$v$}_j$, the left and right singular value basis vectors. These vectors are the set of eigenvectors of the square matrices $\mbox{\boldmath$FF$}^T$ ($N \times N$) and $\mbox{\boldmath$F$}^T\mbox{\boldmath$F$}$ ($M \times M$).

It is straightforward to write the solution to the system defined by Equation (\ref{minLS}) using the SVD in the absence of regularization when $\lambda=0$ in Equation (\ref{minLS}), expressing the solution as a sum over the basis vectors $\mbox{\boldmath$v$}_j$:
\begin{equation}
\mbox{\boldmath$s$}=(\mbox{\boldmath$F$}^T\mbox{\boldmath$F$})^{-1}\mbox{\boldmath$F$}^T
\hat{\mbox{\boldmath$d$}}=\mbox{\boldmath$V \Sigma$}^{-1} \mbox{\boldmath$U$}^T \hat{\mbox{\boldmath$d$}} = \sum_j \frac{\mbox{\boldmath$u$}_j^T \hat{\mbox{\boldmath$d$}}}{\nu_j} \mbox{\boldmath$v$}_j.
\label{FinvSVD}
\end{equation}
In this equation, we have written the SVD in terms of sums over the orthogonal columns of $\mbox{\boldmath$U$}$ and $\mbox{\boldmath$V$}$, and the entries of $\mbox{\boldmath$\Sigma$}^{-1}$, which is simply defined as an $M \times N$ diagonal matrix with non-zero elements $\Sigma^{-1}_{jj}=1/\nu_j$. The SVD allows us to express $\mbox{\boldmath$s$}$ as an expansion over the orthogonal basis $\mbox{\boldmath$v$}_j$.

The matrix $\mbox{\boldmath$F$}$ will have small singular values such that $\nu_j \rightarrow 0$ if the problem is ill-posed. These vanishingly small singular values cause the corresponding terms in Equation (\ref{FinvSVD}) to become large. The solution $\mbox{\boldmath$s$}$ may then become corrupted by the noise contained in the data vector $\hat{\mbox{\boldmath$d$}}$. This amplification of noise due to small singular values is the reason why regularization is required in Equation (\ref{minLS}).

The simplest regularization scheme simply truncates the terms that arise from small singular values from the sum in Equation (\ref{FinvSVD}). Since the singular values form a non-increasing set, this corresponds to discarding all terms $j \ge k$, where $k$ is the truncation threshold. Early termination of the sum removes the high frequency components of the basis vectors $\mbox{\boldmath$v$}_j$. This is known as the truncated singular value decomposition, or TSVD:
\begin{equation}
\mbox{\boldmath$s$}_{\phi}=\sum_j \phi_j \frac{\mbox{\boldmath$u$}_j^T \hat{\mbox{\boldmath$d$}}}{\nu_j} \mbox{\boldmath$v$}_j,
\label{sTik}
\end{equation}
where $\phi_j$ are a set of constants called the filter factors that are equal to $1$ for terms $j \leq k$ and $0$ for all terms higher than this threshold. However, terminating the summation abruptly may discard too much high frequency information. A more general choice is to gradually decrease the contribution of small singular value terms to the sum. This approach is called Tikhonov regularization, which amounts to a modification of the filter factors \citep{tikhonov}:
\begin{equation}
\phi_j=\frac{\nu_j^2}{\nu_j^2+\lambda}
\end{equation}
where $\lambda$ is the regularization constant.
Note that $\phi_j \approx 1$ when $\nu^2_j \gg \lambda$, which occurs for small $j$.  When $\nu_j$ is smaller than the regularization constant (large $j$), the filter factors damp the corresponding terms of Equation (\ref{FinvSVD}) as $\phi_j \approx \nu_j^2 / \lambda$. Thus, $\lambda$ must be assigned a value between the maximum and minimum singular values $\nu_1$ and $\nu_N$. This regularization scheme corresponds to setting the matrix $\mbox{\boldmath$H$}=\mbox{\boldmath$I$}$ in Equation (\ref{minLS}) \citep{twomey, tikhonov}. Regularization modifies the system that we are attempting to solve so that the inverse of Equation (\ref{eqF}) becomes
\begin{equation}
\mbox{\boldmath$F$}_{\phi}^{-1}=(\mbox{\boldmath$F$}^T\mbox{\boldmath$F$}+\lambda \mbox{\boldmath$I$})^{-1}
\mbox{\boldmath$F$}^T=\mbox{\boldmath$V \Sigma$}^{-1} \mbox{ \boldmath$\Phi U$}^T,
\label{regMat}
\end{equation}
where $\mbox{\boldmath$\Phi$}$ is the $N \times N$ diagonal matrix of filter factors with diagonal elements $\phi_{i>M}=0$.

Note that these schemes do not specify how much regularization should be included for a given problem.  The strength of the regularizing effect in Tikhonov regularization is controlled by the value of the regularization constant $\lambda$ and by the truncation index $k$ in the TSVD scheme. The regularization constant is a ``hyper-parameter'' which must be selected a priori. Fortunately, several methods exist to estimate the optimal regularization parameter for a given problem \citep{hansen2}.

\subsection{Regularization Parameter Selection Methods}
\label{secSelectionMethods}

A widely used technique to select a regularization parameter is the L-curve criterion \citep{hansenLcurve}, which we used in \citet{rogersFiege2011}. The L-curve is a plot of the residual versus the regularization term that appears in Equation (\ref{minLS}), and is named for the characteristic shape of the resulting curve. The L-curve is parameterized by the regularization constant $\lambda$ and the position on the plot with the largest curvature represents a balance between the image $\chi^2$ and regularization term \citep{press2007}. This does not imply that an optimally regularized solution has a reduced $\chi^2$ exactly equal to $1$, but should trade-off between the amount of source structure and the quality of the fit.

As an alternative to the L-curve, another well-known regularization selection method is generalized cross validation \citep[GCV;][]{golubHeathWahba}. This is a statistical method that aims to minimize the mean square error, $||\mbox{\boldmath$Fs$}_{\phi} - \mbox{\boldmath$d$}||$, where $\mbox{\boldmath$s$}_{\phi}$ is the optimally regularized solution. We now define the GCV function:
\begin{equation}
G(\lambda)=\frac{||\mbox{\boldmath$d$}-\mbox{\boldmath$Fs$}||^2}{{\rm trace}(\mbox{\boldmath$I$}_N - \mbox{\boldmath$FF$}_{\phi}^{-1})^2},
\label{GCVdefn}
\end{equation}
where $\mbox{\boldmath$I$}_N$ is the $N \times N$ identity matrix. This equation is based on statistical arguments that consider a solution to be properly regularized when it can predict elements of the data vector that have been omitted \citep{hansen1}. The trace term in the denominator can be dramatically simplified given the definition of $\mbox{\boldmath$F$}_{\phi}^{-1}$ in terms of the SVD (Equation (\ref{regMat})). The denominator of the GCV function becomes:
\begin{equation}
{\rm trace} (\mbox{\boldmath$I$}_N -\mbox{\boldmath$FV \Sigma$}^{-1} \mbox{   \boldmath$\Phi U$}^T) = {\rm trace} (\mbox{\boldmath$I$}_N - \mbox{\boldmath$U \Phi U$}^T)
\end{equation}
using the SVD expansion of $\mbox{\boldmath$F$}$ (Equation (\ref{eqF})). With the orthogonality of $\mbox{\boldmath$U$}$ and the diagonality of $\mbox{\boldmath$\Phi$}$, the trace term simplifies dramatically. We are left with ${\rm trace}(\mbox{\boldmath$I$}_N - \mbox{\boldmath$\Phi$})$ such that
\begin{equation}
{\rm trace}(\mbox{\boldmath$I$}_N-\mbox{\boldmath$FF$}_{\phi}^{-1})=N-\sum_i \phi_i.
\end{equation}
This sum represents the number of degrees of freedom in the problem. Putting these arguments together, the GCV function becomes
\begin{equation}
G(\lambda)=\frac{||\mbox{\boldmath$Fs$}_{\phi}-\mbox{\boldmath$d$}||^2}{(N-\sum_i \phi_i)^2}.
\label{GCVF}
\end{equation}
\citet{wahba} showed that when the errors in the data vector are unbiased white noise with covariance matrix $\mbox{\boldmath$C$}=\sigma^2 \mbox{\boldmath$I$}_N$, and satisfy the discrete Picard condition \citep{picard1, picard2}, the minimum of the GCV function corresponds to a regularization parameter that is a good estimator of the optimal $\lambda$ and approaches this value asymptotically as $N \rightarrow \infty$. The convergence results between the true solution of a test problem and the GCV-regularized solution have also been thoroughly explored when these conditions are not satisfied \citep{vogel, lukas}.

The denominator of the GCV function has special significance for gravitational lens modeling. Lens modeling schemes that pixelate the source plane have been criticized for relying on regularization since smoothing causes the number of degrees of freedom in the source to become undetermined \citep{kochanek}. \citet{suyu06} give an estimate for the number of effective degrees of freedom based on Bayesian arguments. In that work the authors construct a variety of possible expressions for the number of degrees of freedom (NDF), and chose  ${\rm NDF}= N-\gamma$ with $N$ the number of image pixels, and
\begin{equation}
\gamma =\sum_{i=1}^{M} \frac{\nu_i^2}{\nu_i^2 + \lambda},
\label{gam1}
\end{equation}
which corresponds to Tikhonov (zeroth order) regularization when $\mbox{\boldmath$H$}=\mbox{\boldmath$I$}$ in Equation (\ref{minLS}). This expression was selected as the correct number of degrees of freedom based on an empirical test that produced a reduced $\chi^2$ nearest to $1$ for a simulated test problem \citep[see Table $1$,][]{suyu06}. In fact, $\gamma$ is simply the sum of the filter factors from Tikhonov regularization. The GCV function gives a statistical argument for choosing this value based on the nature of an optimally regularized source inversion.

In addition to the GCV method, the unbiased predictive risk estimator \citep[UPRE;][]{mallows73} has also been used to select the regularization parameter in deconvolution problems \citep{vogel02, bardsley, lin10}. The UPRE method was initially developed for model selection in linear regression, though variations of this approach have been subsequently applied to the solution of inverse problems. A concise derivation of the method can be found in \citet{vogel02}; here we simply define the UPRE function
\begin{equation}
U(\lambda)=||\mbox{\boldmath$Fs$}_{\phi}-\mbox{\boldmath$d$}||^2+2{\rm trace}\left( \mbox{\boldmath$FF$}_{\phi}^{-1} \right) - N.
\label{UPREdef}
\end{equation}
In analogy with the denominator of the GCV function, we identify the trace term with the sum of the filter factors, $\gamma$. The optimal regularization parameter is chosen as the value of $\lambda$ that minimizes $U(\lambda)$.

Iterative methods complicate the calculation of the GCV and UPRE functions since we do not know the filter factors a priori, nor do we have the decomposition of $\mbox{\boldmath$F$}$, which can be expensive due to the sparsity and size of the matrix. In this case, we estimate the denominator by a Monte Carlo method \citep{girard}. This allows two advantages: we approximate the number of source degrees of freedom while simultaneously finding an approximation to the optimal regularization parameter. Using an iterative method, we find these quantities as we solve for the source intensity distribution. This is accomplished by running iterations on both $\hat{\mbox{\boldmath$d$}}$ and $\tilde{\mbox{\boldmath$d$}}$, where the vector $\tilde{\mbox{\boldmath$d$}}$ is a discrete white-noise vector composed of elements that are $\pm 1$ with equal probability, as commonly used in the image processing literature \citep{hutch,vogel02, bardsley, favati}.

We form the product $\tilde{\mbox{\boldmath$d$}}^T\tilde{\mbox{\boldmath$r$}}$, where $\tilde{\mbox{\boldmath$r$}}=\tilde{\mbox{\boldmath$d$}}-{\mbox{\boldmath$F$}}\tilde{{\mbox{\boldmath$s$}}}_{\phi}$. This quantity approximates the denominator of the GCV function and therefore the number of degrees of freedom in the iterative problem \citep{girard, hansen1}. This calculation requires twice the work during the iterative process and therefore effectively doubles the execution time of the code to solve for the source intensity function. However, since we generally need only a small number of iterations to solve a gravitational lens system, this extra work is acceptable due to the amount of information the calculation provides. By using this Monte Carlo estimate, we find the number of effective degrees of freedom and evaluate Equation (\ref{GCVdefn}) at each iteration. Once we have evaluated an arbitrary number of iterations, we find the minimum of the GCV function and select the critical number of iterations necessary to produce an optimally regularized source. The estimate of the number of source degrees of freedom and the residual at this iteration are used to evaluate the reduced $\chi^2$ of the lens model. A similar procedure can be used to evaluate the UPRE function using $\tilde{\mbox{\boldmath$d$}}^T\mbox{\boldmath$F$}\tilde{\mbox{\boldmath$s$}}_{\phi}$
to approximate $\gamma$, the trace term in Equation (\ref{UPREdef}).

\citet{rogersFiege2011} explored the L-curve method for the selection of regularization parameters in gravitational lens modeling, arguing that the L-curve provides a useful parameter selection method that yields results which are easy to interpret. However, using this selection criterion can be difficult due to the curvature calculation, which requires spline fitting of the points on the L-curve and the curvature of the resulting smoothed curve.  This calculation is non-trivial and results can be somewhat sensitive to the details of the fitting procedure.  The GCV and UPRE functions require more involved statistical arguments but provide more robust and reliable selection methods, since the functions are calculated at each iteration simultaneously with the linear optimization. We find that the GCV and UPRE methods produce results that are consistent with one another, indicating that both can be used effectively to determine the optimal termination condition for the iterative solver.  We prefer evaluating the GCV and UPRE functions to the L-curve method for the reasons outlined above and focus on these parameter selection routines in this study.

\section{The SLACS Survey}

The Sloan Lens ACS Survey (SLACS) was conducted using the Hubble Space Telescope ACS instrument \citep{SLACSI}. The survey has detected $70$ early type galaxies with definite lensed sources in the redshift range $z=0.06$ to $z=0.33$. The candidate systems were chosen by spectral analysis of galaxies in the luminous red galaxy (LRG) and MAIN samples of the Sloan Digital Sky Survey (SDSS; http://www.sdss.org). Potential gravitational lens candidates were discovered when two distinct redshifts were seen within a single SDSS spectrum. We use reduced SLACS data from \citet{bandara}, who modeled the surface brightness of the E/S0 lens galaxies using the sum of two components, a S\'{e}rsic bulge (Equation (\ref{sersicEq})) and an exponential disk. The PSF model from the ACS library was used in the surface brightness subtraction, making use of the GIM2D code \citep{simard}. All of the data are F814W $I$-band images. See \citet{bandara} for more details on the reduction procedure.

\section{Results}

\citet{SLACSV} modeled the SLACS gravitational lens systems using analytical S\'{e}rsic and Gaussian source models to describe the intensity distribution in the source plane. A subset of $15$ of these systems were further investigated using the semilinear method \citep{SLACSIII}. We focus on six of the SLACS lens systems in this paper, and plan to model more of them in the future. Since they have been well studied using several established methods, the SLACS galaxies provide a useful consistency check for verifying the results of our lens modeling code.

The SLACS systems are modeled using a singular isothermal elliptical mass density (SIE). We define a distance $\psi=\sqrt{ q x^2 + y^2 / q }$, such that the deflection angle $\mbox{\boldmath$\alpha$} = (\alpha_x, \alpha_y)$ is given by
\begin{equation}
\alpha_x=\frac{b}{q_f}\tan^{-1}\left( \frac{q_f x}{\psi}  \right)
\end{equation}
\begin{equation}
\alpha_y=\frac{b}{q_f}\tanh^{-1}\left( \frac{q_f y}{\psi}  \right),
\end{equation}
with $q_f=\sqrt{1/q - q}$, and Einstein radius $b$. In the limit $q \rightarrow 1$, the model corresponds to a singular isothermal sphere with Einstein radius
\begin{equation}
b = 4 \pi \frac{\sigma_v^2}{c^2}\frac{D_{ds}}{D_{s}},
\end{equation}
where $\sigma_v$ is the velocity dispersion, $c$ the speed of light, $D_{ds}$ the distance between the deflector and the source, and $D_s$ the distance between the observer and the source. These distances depend on the corresponding redshifts $z_d$ and $z_s$ and determine angular diameter distances that depend on the cosmological model used. We assume a standard cosmology with Hubble constant $H_0=70$ km s$^{-1}$ Mpc$^{-1}$, matter density $\Omega_0 = 0.3$ and cosmological constant $\Lambda_0 = 0.7$. Following \citet{SLACSV}, we adopt the intermediate-axis normalization of the SIE \citep{kormann94}. This normalization fixes the mass within given isodensity contours for constant $b$, and is implemented in the deflection angles above. \citet{SLACSIII} showed that the SIE is a useful model of early type isolated galaxies because the lens density ellipticity and orientation were found to align well with the surface brightness of the SLACS lens galaxies, indicating that light closely traces mass for these systems. No significant external shear was found to improve the fits. We therefore follow \citet{SLACSIII} and adopt the SIE as a good model to represent isolated the early type E/S0 SLACS lens galaxies.

We cropped out the residuals left over from the surface brightness subtraction of the lens in the F814W SLACS data, and cropped the field of view to the region of interest, but performed no rebinning or other manipulation of the data in any way. Our lens models use the same ACS PSF that was used for the lens galaxy subtraction. Although it is known that the ACS PSF is position dependent \citep{bandara}, we simplify our treatment by assuming a constant PSF over the region of interest, though we have previously developed methods to include spatially variant PSFs in the gravitational lens problem \citep{rogersFiegePSF}. We output the sigma image from the GALFIT code \citep{GALFIT} that corresponds with the region of interest to estimate the errors on the image plane. We emphasize that the main focus of this work is to study the regularizing properties of the CGLS method on the derived solutions with the GCV and UPRE schemes to select the optimal level of regularization.

Our analysis initially solves for the parameters of an analytical source model, which we use as an approximate solution to a more refined model that uses a pixelated source.  We start by treating the source plane intensity distribution as a sum of S\'{e}rsic profiles, using the same number of analytical source components to model each system as in \citet{SLACSV}. The SIE lens is used to find the lensed image of the source plane, which is convolved with the appropriate ACS PSF. We search for the global minimum of $\chi^2$, using the Ferret GA \citep{qubist} to fit both the lens density and source parameters. Once we find an approximation to the global minimum, we select a volume of lens parameter space in the neighborhood around the best fit lens model.  Noting that Ferret is used predominantly as a bounded optimizer, this neighborhood becomes the search volume in the next step of our method, which replaces our analytical source model with a pixelated source. The optimization of a pixelated model requires a new Ferret run, which begins with the search volume found in the previous step populated initially by random lens models.  Normally, we expect the lowest $\chi^2$ model to reside within this volume; however, we configure the optimizer using ``soft'' boundaries, which allows the GA to move outside of the predefined search volume if the initial approximation is bounded too tightly. This option allows Ferret to expand the search space if a large fraction of the GA population occupies positions close to the boundaries of the parameter space. In general, the lens parameters of our pixelated sources were found to reside within these search volumes and agree well with the analytical approximations. We compute our best refined model by optimizing the lens and source plane parameters using a pixelated source and regularizing iteration selected by the GCV and UPRE functions.

In addition to the regularizing effect of truncated iteration, we have found that enforcing non-negativity in the source solutions dramatically improves the quality of the reconstruction and tends to further reduce remaining structure in the image residuals. As a final step, we have modeled the set of best-fit lens models with the modified residual norm steepest descent algorithm \citep[MRNSD;][]{kaufman, nagyMRNSD, bardsley}. This algorithm is a bounded SD optimization routine that seeks sources with $s_j \geq 0$. In practice we have found that the MRNSD method can produce residuals which decrease in a step-like manner, making the determination of the minimum difficult for the GCV and UPRE functions. This zig-zag behavior has also been noted by \citet{favati} in the context of the standard deconvolution problem.

Combining analytical and pixelated sources greatly improves the efficiency of the search, since analytical models can be evaluated very quickly. Searching using pixelated sources is a more intensive process, and time can be saved by adopting the semilinear method only once we have a good approximation to the lens parameters corresponding to the minimum $\chi^2$. \citet{rogersFiege2011} noted that a set of trivial pixelated solutions exist when global optimization methods are used to model lensed systems. These trivial solutions are found when the effect of the lens is reduced, resulting in sources that closely resemble the data. The two-stage optimization process is useful since the initial analytical sources are generally not as flexible as pixelated sources, and thus provide a natural method for avoiding exploration of the trivial regions of the parameter space. The analytical stage of the algorithm terminates once the GA has converged and we no longer see improvement in the population. Typically, convergence requires only $50 - 100$ generations using a population of $300$ individuals for the analytical portion of the optimization, and approximately $100$ iterations for the second semilinear optimization stage.

\begin{figure}
\scriptsize
\hspace*{-2.15cm}
\epsscale{1.25} \plotone{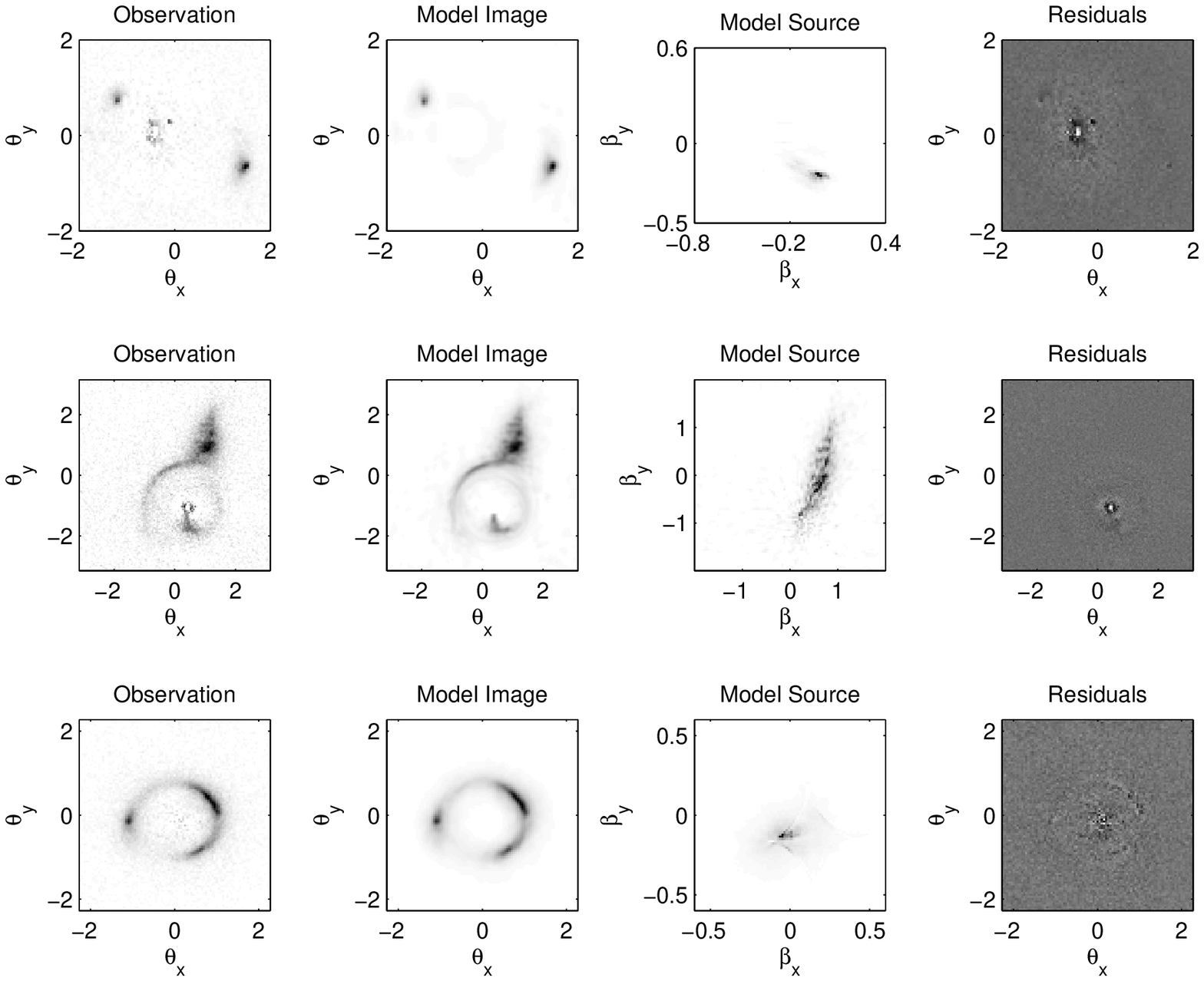}
\caption[A Variety of SLACS Lens Models, I]
{\label{bigA} A selection of SLACS gravitational lenses. The sources are non-negative and found using the MRNSD algorithm as the final polishing step. The columns show the data $\mbox{\boldmath$d$}$, image model, source model $\mbox{\boldmath$s$}$ and residual $\mbox{\boldmath$r$}$ respectively. The model parameters are given in table \ref{table1}. Top row: SDSS J0037-0942, second row: SDSS J0216-0813, bottom row: SDSS J0737+3216. Model images and sources in this figure were produced using an image pixel subsampling factor of $3$.}
\end{figure}

\begin{figure}
\scriptsize
\hspace*{-2.15cm}
\epsscale{1.25} \plotone{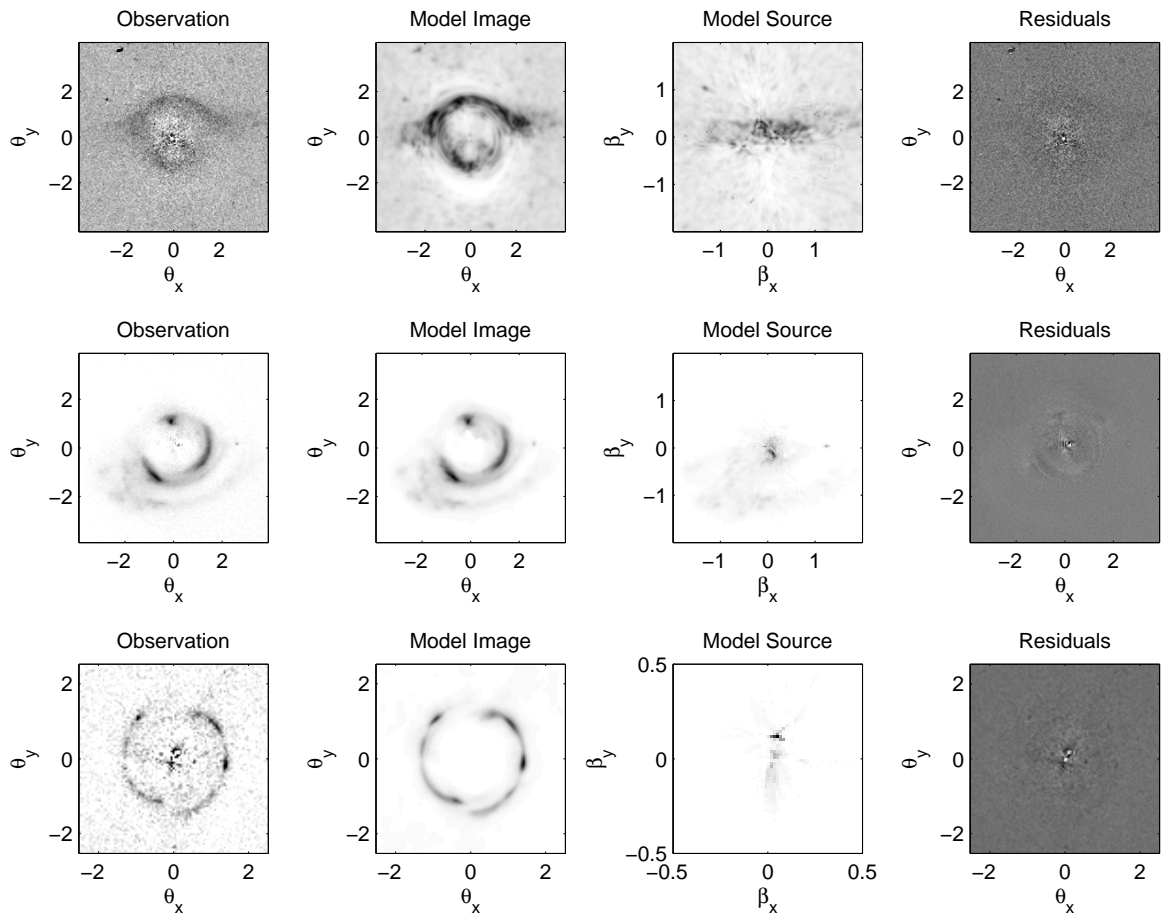}
\caption[A Variety of SLACS Lens Models, II]
{\label{bigB} Top row: SDSS J0912+0029, second row: SDSS J0956+5100, bottom row: SDSS 1402+6321. Model images and sources in this figure were produced using an image pixel subsampling factor of $3$.}
\end{figure}

The final velocity dispersion $\sigma_v$, axis ratio $q$, and Einstein radius $b$ of our models are shown in Table \ref{table1}. The reduced data, model image, recovered non-negative source and residuals are shown in Figures \ref{bigA} and \ref{bigB}.
\begin{table}
\begin{center}
\begin{tabular}{|l|l|l|l|l|l|}
\hline
\multicolumn{6}{|c|}{Table 1 - Lens Model Parameters} \\
\hline
SDSS System & $z_d$       & $z_s$       & $\sigma_v$ (km s$^{-1}$) & $q$ & b('')  \\ \hline
J0037-0942   & $0.1955$ & $0.6322$ & $286$ & $0.825$ & $1.55$ \\ \hline
J0216-0813   & $0.3317$ & $0.5235$ & $351$ & $0.783$ & $1.18$ \\ \hline
J0737+3216  & $0.3223$ & $0.5812$ & $291$ & $0.661$ & $0.99$ \\ \hline
J0912+0029  & $0.3240$ & $0.1642$ & $341$ & $0.561$ & $1.59$ \\ \hline
J0956+5100  & $0.2405$ & $0.4700$ & $318$ & $0.620$ & $1.33$ \\ \hline
J1402+6321  & $0.2046$ & $0.4814$ & $292$ & $0.843$ & $1.34$ \\ \hline
\end{tabular}
\caption[SLACS Lens Modeling Results]
{\label{table1} Lens model parameters for a subset of the SLACS systems found by the Ferret GA with source reconstruction by the CGLS routine.}
\end{center}
\end{table}
Our results agree with the SLACS lens models for each system to within $3 \%$ in velocity dispersion $\sigma_v$. Both the pixelated and analytical source plane intensity distributions agree with one another in all cases. Our lens modeling results agree with the parameters in \citet{SLACSV} very well. The reduced $\chi^2$ statistic for all systems is very close to unity.

\begin{figure}
\scriptsize
\hspace*{-2.0cm}
\epsscale{1.0} \plotone{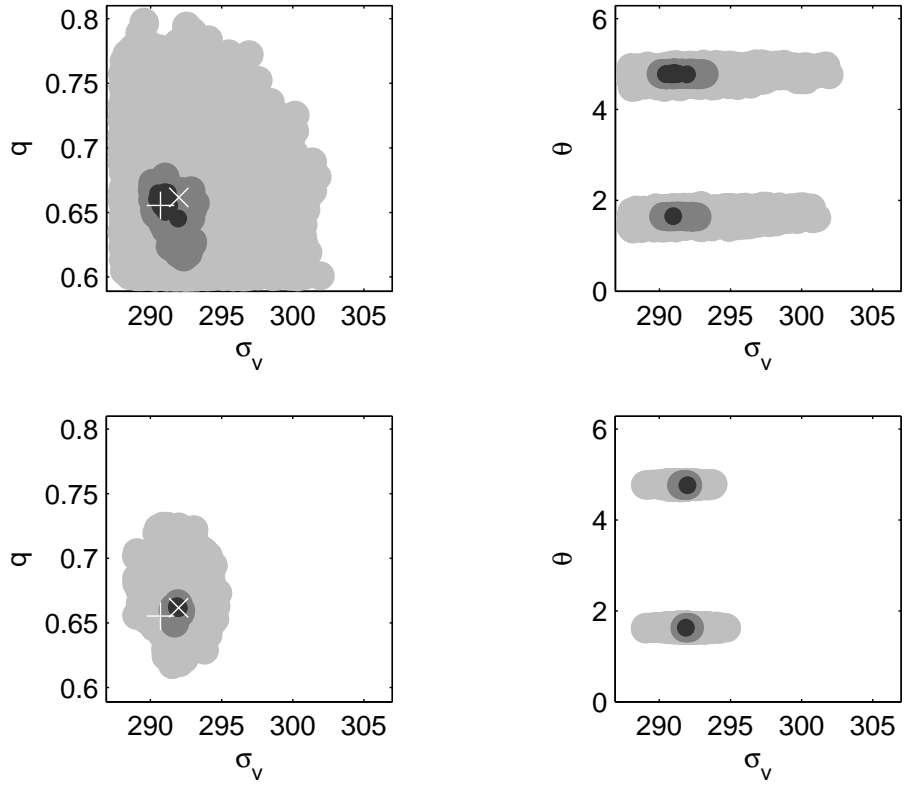}
\caption[Parameter space mapping of SDSS J0737+3216]
{\label{parameterSpace} An example of the lens parameter space mapping for J0737+3216 using the GA. Solutions within the $1 \sigma$ contour are black, $2 \sigma$ mid gray and $3\sigma$ light gray. The top row shows the lens parameter space with a pixelated source (lens velocity dispersion $\sigma_v$, ellipticity $q$ and lens orientation angle $\theta$), and the bottom row shows the lens parameter space for the analytical source. Note that the position of the best fit lens model changes when a pixelated source is used. The right column demonstrates the expected degeneracy between the velocity dispersion and lens orientation angle.}
\end{figure}

The Einstein radius (velocity dispersion) and ellipticity of SDSS J$0737+3216$ are similar to the model from \citet{SLACSV}, and share a velocity dispersion similar to \citet{marshall}. However, the recovered ellipticity is smaller than both \citet{SLACSIII} and \citet{marshall}. We found a lower ellipticity from both the initial and analytical source fit and by pixelated source modeling. To illustrate the difference between analytical and pixelated source lens models, we show the lens parameter space in Figure \ref{parameterSpace}. Points in this figure are shaded according to confidence interval, and demonstrates that the $1 \sigma$, $2 \sigma$ and $3 \sigma$ confidence regions are larger for the pixelated source than the analytical source. This shows that a more flexible pixelated source can broaden the error bars on the lens parameters when compared to an analytical description. We have observed this result for all the lens systems that we studied. The SDSS J$0912+0029$ data is heavily contaminated with noise, although it is adequately fit by our GCV and UPRE regularized solution, and our analytical and pixelated sources agree. Of all of the systems, SDSS J$0956+5100$ and SDSS J$0737+3216$ show the most structure in the residuals, although the magnitude of these residuals are small ($<1 \%$) compared to the intensities of the image pixels. In fact, the largest systematic effects present in most of the residual images in Figures \ref{bigA} and \ref{bigB} are produced from the subtraction of the intensity profile of the lens galaxy. The GCV and UPRE selected iterations are identical for the lens parameters above, and have been observed to generally differ by only a few iterations.

Overall we are encouraged by our results since we were able to recover the SLACS lens parameters and general source morphologies. The results could be improved slightly by including a final local optimization step to `polish' the results returned from the GA. We did not detect any parameter space degeneracies except for the expected position angle degeneracy that leaves the solution unchanged when the elliptical mass distribution is rotated by $180^{\circ}$.

We have used the GCV and UPRE approaches with both CGLS and SD, and find similar results for both of these algorithms. The SD routine takes much longer than the CGLS method to converge, although it is in general a more stable approach to regularization and has been suggested as a superior routine for image deblurring problems due to its reduced sensitivity to stopping criterion \citep{nagyPalmer}. The best fit MRNSD solutions are found by comparing the solution at each iteration to the optimally regularized CGLS solution. This comparison minimizes
\begin{equation}
z=\frac{||x^{ {\rm CGLS} }-x^{{\rm MRNSD}}_k||}{||x^{ {\rm CGLS}}||}
\label{solutionComp}
\end{equation}
where $x^{ {\rm CGLS} }$ is the optimally regularized CGLS solution and $x^{{\rm MRNSD}}_k$ the non-negative solution at the $k$th iteration of the MRNSD algorithm. In general the MRNSD residuals appear smoother than the residuals of the CGLS models. This is due to the reconstruction of back-traced noise present in the CGLS solutions. The filter factors of the CGLS method are given by a recursion relation that depends on all of the singular values \citep{hansen2}. Even though CGLS tends to suppress high frequency noise at the beginning of the optimization process, the high frequency components are not completely damped out at any given iteration and build up over the course of a run. Hence, even the optimally regularized solution still contains some high-frequency components that correspond to back-traced noise. The MRNSD algorithm seems to be more robust to the propagation of high-frequency noise in the recovered non-negative solutions, thus producing images that are naturally smoother than the corresponding CGLS sources.

Regularization by truncated iteration in the context of Krylov optimization is the simplest of many regularization methods that can be used. Truncated iteration regularization produces solutions (figures \ref{bigA} and \ref{bigB}) which are less smooth than the second order (curvature) regularization used in \citet{SLACSIII}. It has been suggested that the LSQR algorithm \citep{bjorck} can generally accomfffmodate more complicated regularization schemes with increased numerical stability when the system is poorly conditioned. The LSQR routine is an iterative Krylov subspace method that solves least-squares problems using QR decomposition \citep{paigeSaunders}. We previously tested LSQR in the context of gravitational lens modeling using simulated data with the L-curve \citep{rogersFiege2011}, and suggest that this scheme may provide a higher level of control over regularizing effects than the truncated iteration scheme used in this study.

\begin{figure}
\scriptsize
\hspace*{-2.5cm}
\epsscale{1.25}\plotone{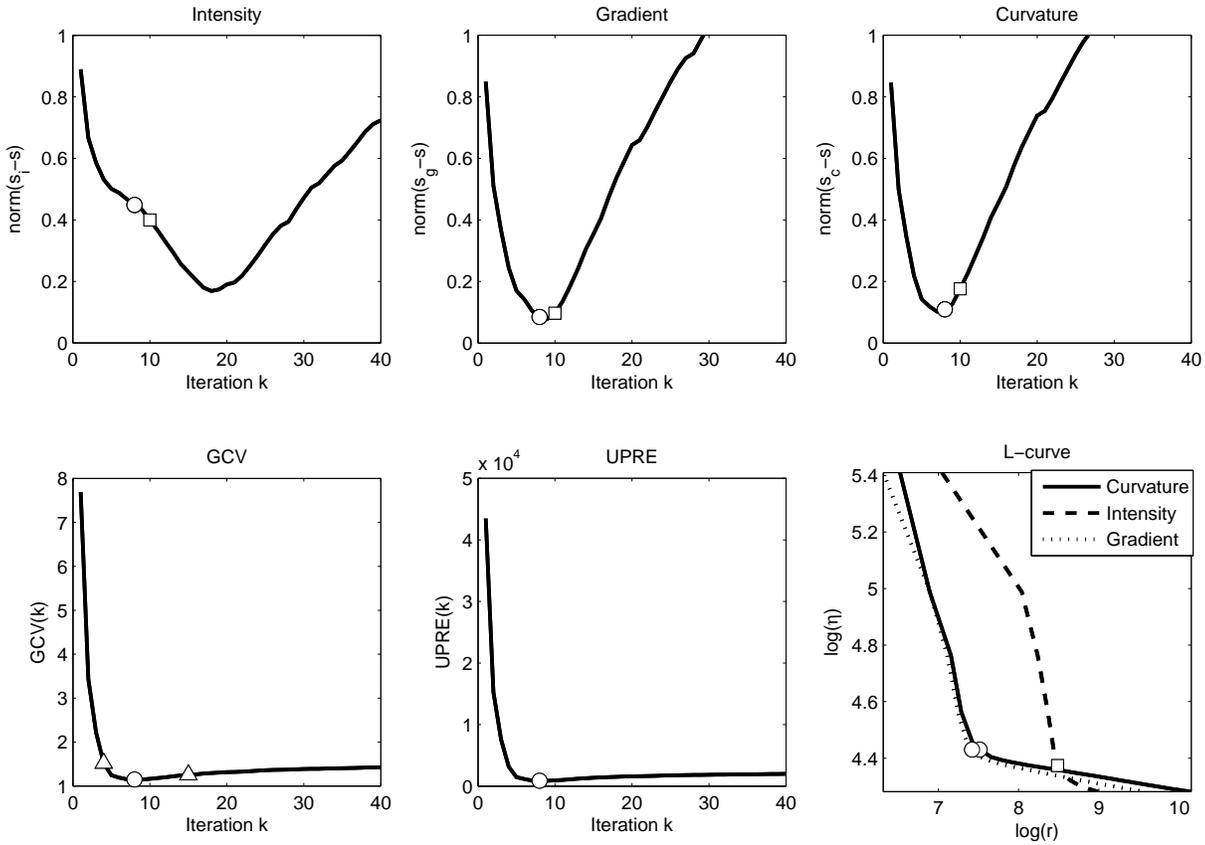}
\caption[A comparison of regularization schemes]
{\label{rev_compare} Top row: Comparison of the difference between CGLS at each iteration $k$ and Bayesian selected solutions using intensity (top left panel), gradient (top center) and curvature regularization (top right). Early iterations correspond more closely to gradient and curvature regularized solutions while the later iterations more closely approximate intensity regularization. The GCV and UPRE functions select the iteration marked by circles, and the intensity L-curve picks the under-regularized solution marked with a square. Bottom row: Selection methods as a function of iteration including the GCV function (bottom left panel), the UPRE function (bottom center) and a set of L-curves found by using the intensity, gradient and curvature norms of the CGLS solutions. In this case the curvature and gradient L-curves select the same solution as the GCV and UPRE functions. The iterations marked with triangles on the GCV plot represent over-regularized and under-regularized solutions, respectively.}
\end{figure}

Figure \ref{rev_compare} illustrates the regularizing behavior of the CGLS routine as a function of iteration $k$ using SDSS J$0216-0813$ as an example. Iterative methods like CGLS produce a sequence of solutions, which we compare with the optimally regularized Bayesian solutions found by the method of \citet{suyu06}, making use of Equation (\ref{solutionComp}). The top panels of this figure compare the solution at each iteration with intensity (zeroth order), gradient (first order), and curvature (second order) regularization used with the semilinear method. The solution at earlier iterations is more heavily regularized, with a larger portion of high frequency components damped. These early iterations produce solutions that resemble the results of the semilinear method using gradient and curvature regularization terms. The solution from later iterations contains a larger number of high frequency components, which simulates the effect of using zeroth order regularization in the semilinear method. Note that the truncated iterations of the CGLS method do not produce identical solutions to those found from the semilinear method, which is not surprising since the filter factors of the CGLS approach differ significantly from linear regularization methods.

The GCV and UPRE functions are plotted on the lower left and center panels of Figure \ref{rev_compare}. The selected stopping iteration is found from the location of the minima of these functions, which correspond with one another in all of our test cases in Table \ref{table1} and are marked with circles. These selection schemes both favor the solutions from early iterations. The lower right panel plots the L-curves using all three types of regularization. Note that in this case the L-curves from gradient and curvature regularization match the GCV and UPRE solutions. However, we often observe that the L-curve can show false curvature maxima when iterative optimizers make rapid progress early in the run, leading to dramatically over regularized solutions. The GCV and UPRE functions avoid this problem. Combined with the statistical arguments used to derive the GCV and UPRE functions and the more robust behavior of these regularization parameter selection methods, we conclude that the GCV and UPRE approaches are more useful than the L-curve methodology for solving the least-squares source deconvolution problem for gravitational lens systems. The comparison of our results with the optimally regularized Bayesian solutions is shown in Table \ref{rev_table} for each of the systems that we modeled.

\begin{table}
\begin{center}
\begin{tabular}{|l|l|l|l|l|l|l|}
\hline
System & ${\rm N}_{\rm img}$ & ${\rm N}_{\rm src}$ & Reg. & ${\rm N}_{\rm eff}$ & $\gamma$ & $\chi^2$ \\
\hline \hline

\multirow{4}{*}{J0037-0942} & \multirow{4}{*}{1072} & \multirow{4}{*}{895} & I & 708.90 & 363.10 & 0.94\\
 & & & G & $799.35$ & $272.65$ & $1.03$ \\
 & & & C & $845.29$ & $226.71$ & $1.08$ \\
 & & & 12 Iter. & $738.53$ & $334.47$ & $0.97$ \\
\hline \hline

\multirow{4}{*}{J0216-0813} & \multirow{4}{*}{6599} & \multirow{4}{*}{2158} & I & $5524.13$ & $1074.87$ & $1.01$ \\
 & & & G & $6074.54$ & $524.46$ & $1.05$ \\
 & & & C & $6247.78$ & $351.22$ & $1.07$ \\
 & & & 8 Iter. & $6130.83$ & $468.17$ & $1.06$ \\
\hline \hline

\multirow{4}{*}{J0737+3216} & \multirow{4}{*}{2536} & \multirow{4}{*}{1217} & I & $1983.95$ & $552.05$ & $0.97$ \\
 & & & G & $2154.11$ & $381.89$ & $1.03$ \\
 & & & C & $2191.58$ & $344.42$ & $1.08$ \\
 & & & 10 Iter. & $2133.33$ & $402.67$ & $1.06$ \\
\hline \hline

\multirow{4}{*}{J0912+0029} & \multirow{4}{*}{9870} & \multirow{4}{*}{2500} & I & $9192.58$ & $677.41$ & $0.96$ \\
 & & & G &  $9659.28$ & $210.72$ & $1.00$ \\
 & & & C & $9764.23$ & $105.77$ & $1.01$ \\
 & & & 5 Iter. & $9482.82$ & $387.18$ & $0.99$ \\
\hline \hline

\multirow{4}{*}{J0956+5100} & \multirow{4}{*}{4622} & \multirow{4}{*}{900} & I &  $3982.96$ & $639.04$ & $0.94$ \\
 & & & G &  $4228.21$ & $393.79$ & $0.98$ \\
 & & & C & $4308.46$ & $313.54$ & $1.01$ \\
 & & & 12 Iter. & $4200.65$ & $421.35$ & $1.01$ \\
\hline \hline

\multirow{4}{*}{J1402+6321} & \multirow{4}{*}{3398} & \multirow{4}{*}{2940} & I &  $2955.95$ & $443.05$ & $0.92$ \\
 & & & G &  $3102.21$ & $295.79$ & $1.00$ \\
 & & & C & $3183.88$ & $214.12$ & $1.06$ \\
 & & & 7 Iter. & $3132.38$ & $265.62$ & $1.03$ \\
\hline

\end{tabular}
\label{table}
\caption[Comparison with the semilinear method using Bayesian regularization parameter selection]
{\label{rev_table} A comparison of Bayesian selected regularization using the semilinear method with the GCV and UPRE functions. ${\rm N}_{\rm img}$ and ${\rm N}_{\rm src}$ are the number of pixels in the image plane and source plane. The column marked ``Reg.'' is the regularization type for each system (I, Intensity; G, gradient; C, curvature), ${\rm N}_{\rm eff}$ is the effective number of degrees of freedom for the corresponding lens model in Table \ref{table1}, and $\gamma$ gives the effective number of source degrees of freedom. The reduced $\chi^2$ is given in the rightmost column for each method. The GCV and UPRE functions gave identical results for each of the systems using the CGLS method. For these calculations we estimated ${\rm N}_{\rm eff}$ and $\gamma$ using the Monte Carlo approach described in Section \ref{secSelectionMethods}.}
\end{center}
\end{table}

\begin{figure}
\scriptsize
\hspace*{-2.0cm}
\epsscale{1.0} \plotone{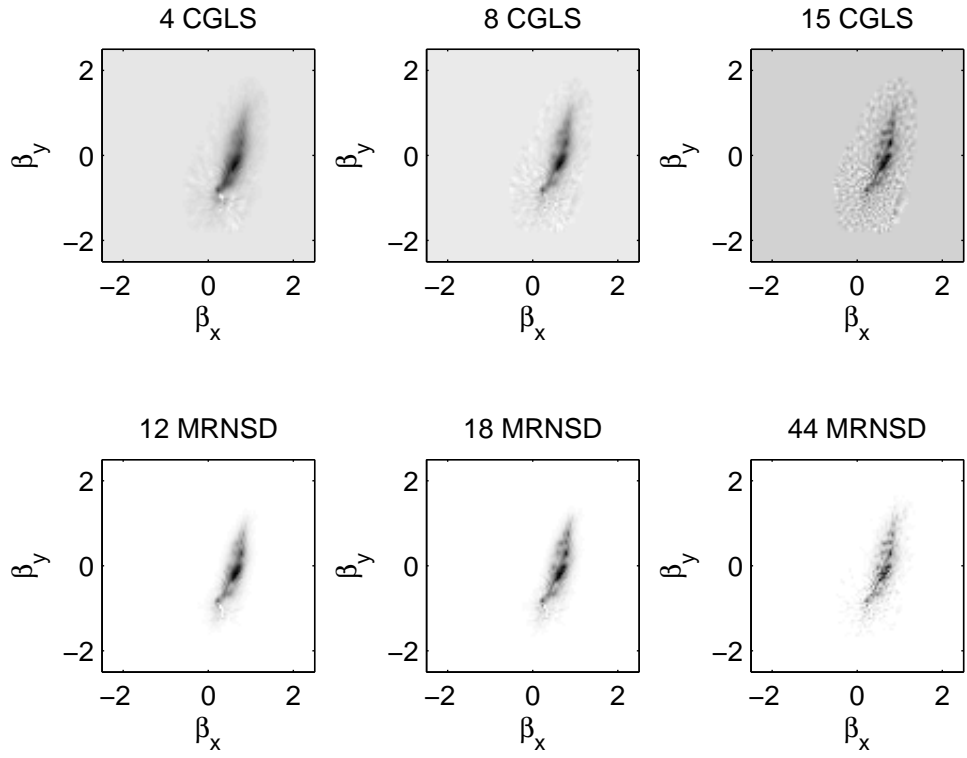}
\caption[An Example of Regularization Effects on a SLACS Lens]
{\label{regJ0216} Three solutions for SDSS J0216-0813 marked in the lower left-hand panel of Figure \ref{rev_compare}. These solutions correspond to over-regularized (left), critically-regularized (middle) and under-regularized solutions (right) as selected by the GCV function. Note the emphasis on back-traced noise in the under-regularized CGLS solution and the excessive smoothing of the over-regularized solution. Non-negative MRNSD solutions are shown on the second row.}
\end{figure}

We have marked two additional points on the GCV curve in Figure \ref{rev_compare}. These points signify over regularized and under regularized solutions. The sources corresponding to the solution of the SDSS J$0216-0813$ system at these iterations are shown in Figure \ref{regJ0216} using both CGLS and MRNSD algorithms. As shown in this figure, the over-regularized solutions are over-smoothed, and the under-regularized solutions include too many high frequency components. The corresponding MRNSD solutions were found by terminating iterations when Equation (\ref{solutionComp}) is minimized. Furthermore, note the presence of back-traced noise in the reconstructions using CGLS, while this noise is effectively suppressed in the MRNSD solutions.

In general, the difficulty in making use of any regularization parameter selection scheme is the need to evaluate each model using a range of regularization constants. This process can be expensive if the time for each evaluation is large. The benefit in making use of an iterative scheme is that each iteration can be thought of as a discrete regularization parameter. Iterative methods are attractive since expensive matrix inverses are not calculated directly, and a regularization parameter selection method can be used at each iteration to determine the optimal regularization strength (i.e., the optimal stopping iteration). Since the GCV and UPRE functions can be evaluated while iterations run, an optimally regularized solution can typically be found in the time needed to solve a system using a single value of the regularization constant with the semilinear method. This time savings is important when using global optimization schemes since a large number of models need to be evaluated with these methods.

Another attractive prospect of iterative methods is the fact that iterations can be carried out without explicit representations of the matrices themselves. For example, subroutines which perform the lensing and blurring operations can be substituted for $\mbox{\boldmath$B$}$ and $\mbox{\boldmath$L$}$ while preserving the least-squares form of the problem. Therefore, even large scale systems that would prohibit the direct application of the semilinear method in matrix form can be practically solved and optimally regularized by an iterative Krylov method \citep{rogersFiege2011, rogersFiegePSF}. For example, \citet{alard} has modeled the cluster lens SL2SJ021408-053532, which is comprised of a small group of $6$ galaxies and results in a set of large lensed arcs. As noted in that work, the large size of the system prevents direct application of the semilinear method. However, iterative approaches with algorithmic substitutions for the explicit representations of the blurring and lensing matrices can accommodate large-scale lensing problems that are realistic for a number of practical modeling situations.

\section{Conclusions}
We have used iterative Krylov methods to model a subset of the SLACS lenses using GCV and the UPRE to select the optimal regularizing iteration. We addressed the problem of the number of effective degrees of freedom in the source by making use of parameter choice methods that are commonly used in standard image deconvolution problems. This approach leads to a key result from \citet{suyu06} that was derived using Bayesian methods. The GCV and UPRE functions shed light on the concept of optimally regularized sources and provide an efficient method to select regularization parameters for iterative schemes. A non-negative bounded iterative algorithm is found to significantly improve the quality of the reconstructed sources. This approach provides non-negative solutions through linear optimization, which is significantly simpler to implement than other constrained optimization techniques such as the maximum entropy method \citep{SkillingBryan, LMEMwayth} that require the use of more complicated non-linear optimization schemes.

The lens parameters recovered by the Ferret GA are similar to previously published results found by \citet{SLACSV} and we find consistency between analytical approximations to the source plane intensity based on a sum of S\'{e}rsic profiles. We plan to investigate a larger sample of the SLACS lenses in the future and explore other local optimization methods to solve the least-squares problem with a variety of regularization schemes.

\section{Acknowledgements}
The authors are grateful to Kaushala Bandara, who supplied the reduced SLACS observations used in this study. We would also like to thank the anonymous referee, whose thoughtful comments and suggestions significantly expanded the scope of this work and
improved the overall flow of the paper.  A.R. acknowledges NSERC for funding this research, and J.F. acknowledges funding from an NSERC Discovery Grant.

\clearpage
\newpage

\clearpage
\newpage

\clearpage
\newpage

\clearpage
\newpage

\clearpage
\newpage

\end{document}